\documentclass{ws-ijmpd}

\begin{document}

\markboth{Sigismondi}
{Sunsets and Solar Diameter Measurement}

%
\catchline{}{}{}{}{}
%

\title{SUNSETS AND SOLAR DIAMETER MEASUREMENT}

\author{Costantino Sigismondi}

\address{Sapienza University of Rome, Physics Dept. P.le Aldo Moro 5 \\
Roma, 00185, Italy\\
University of Nice-Sophia Antipolis (France); IRSOL, Istituto Ricerche Solari di Locarno (Switzerland)\\
sigismondi@icra.it}

\maketitle

\begin{history}
\received{10 Jun 2011}
\revised{Day Month Year}
\comby{Managing Editor}
\end{history}

\begin{abstract}
A sunset over the sea surface offers the possibility to chronometrate a solar transit across the horizon.
The  vertical solar diameter is proportional to the duration of the sunset, the cosine of the azimuth 
and the cosine of the latitude of the observing site. The same formula applies to every circle of equal height, called in arabic almucantarat,
and it is exploited in the measurements of the solar diameter made with the Danjon's solar astrolabes.
The analogies between sunsets and astrolabes observations are presented, showing advantages and sources of errors of these methods of solar astrometry. 

\end{abstract}


\section{Introduction: Method and Rationale}
The method of measurement of the solar diameter with the timing of a transit across a circle of a given altitude,
called $almucantarat$ using a classical arabic word, can be applied also when that circle is the visible horizon, during a sunset,
or sunrise.

The degradation of the image quality, as the Sun approaches the horizon, because of the atmospheric turbulence and because of
the reduced transparency of the atmosphere, makes this measurement rather complicate.

The random nature of the turbulent phenomena occurring
along the line of sight produces stable observational conditions 
during the whole time between the first and the last contact
of the solar disk with the sea horizon.

This stability recalls the advantages of the method of the drift-scan, where the optical defects present along the line of sight act as systematic errors. We assume that they are the same for the first and the last contact of the solar disk with the sea horizon, and the accuracy of the method of diameter measurements depends only on the timing accuracy. 

The departures of this ideal situation are here discussed.

The analysis of the sources of error in the sea sunset measurements of solar diameter helps to present the generalities of the drift-scan method applied to equal almucantarats.
The application of this method to naked-eye and video measurements of sunsets shows the accuracy of the method from an experimental point of view.

\section{Sources of Errors}

\subsection{Black drop}

Let consider the sunset, with the Sun approaching the sea surface. The first contact is difficult to be discerned because of an optical phenomenon traditionally called "black drop". This phenomenon, described firstly for the transit of Venus of 1769 observed by the astronomer Charles Green and the Captain James Cook in Thaiti, is due to the convolution between the point spread function of the telescope (or the eye, or the videocamera) and the solar limb darkening function rapidly varying with respect to the inward distance from the limb.\cite{Pasachoff}

In the case of the sunset the solar limb darkening function is also multiplied by the transmittance of the atmosphere. In Fig. 1 there is the profile of intensity of the solar image near the sunset: on the left there is the lower limb of the Sun, and its luminosity profile is shallower than the higher limb. The altitude of the center of the Sun above the sea horizon is 30 arcmin.  

\begin{figure}
\centerline{\psfig{file=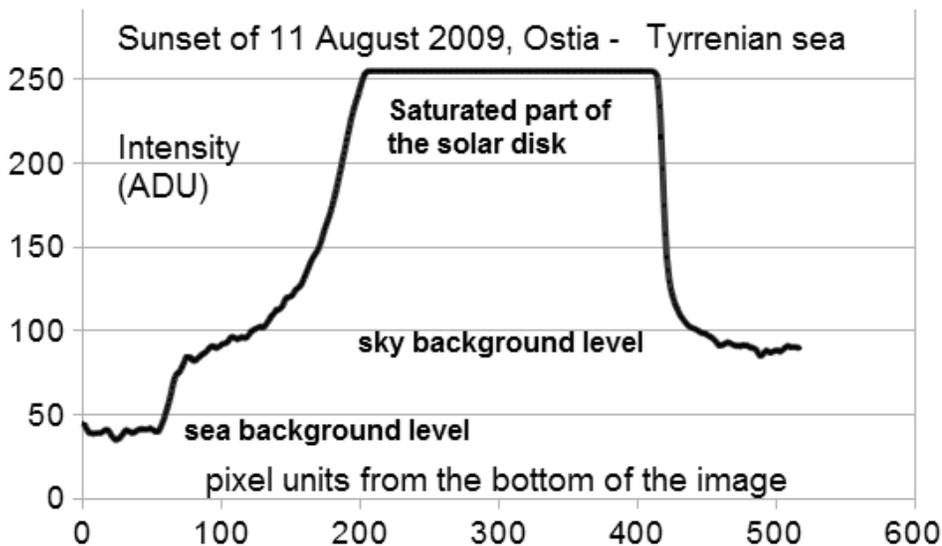,width=13cm}}
\vspace*{0pt}
\caption{Sunset of August 11, 2009: the profile of the intensity of the solar disk, the sky background and the sea background are reported in ADU, Analog-to-Digital Units, or digital counts. The 8 bits sensitivity, 256 level of intensity, has been used.}
\end{figure}

The asymmetry is due to the transmittance of the atmosphere in the lower layers above the sea. At the Tyrrenian sea, during my observations, one of them is reported in Fig. 2, the effect of the lower layers of humidity has been parametrized in 4 additional airmasses at the horizon, with 3 extra-airmasses at 45 arcmin above the horizon (corresponding to the upper limb of the Sun in Fig. 1) and 3.8 extra-airmasses at 15 arcminin above the visible horizon (corresponding to the lower - left limb of Fig.1). 
The intensity of the solar disk,  has been measured relatively to the sky background with the following method. A pinhole has been used to produce an image of the Sun, projected on a white screen. The screen is illuminated either by the pinhole and by the sky background. The distance $D_{max}$ where the contrast of solar image no longer permits its observation is where the intensity of the projected image is equivalent to the reflected sky background on the same screen: $B_{sky}=I_{0}/D_{max}^2$, where $I_{0}$ is the light gathered by the pinhole opening. Since the intensity of the projected image is proportional to the inverse square of the distance between the pinhole and the screen,$D_{max}^2\propto I_{0}$ if we consider as constant the sky background. 

\begin{figure}
\centerline{\psfig{file=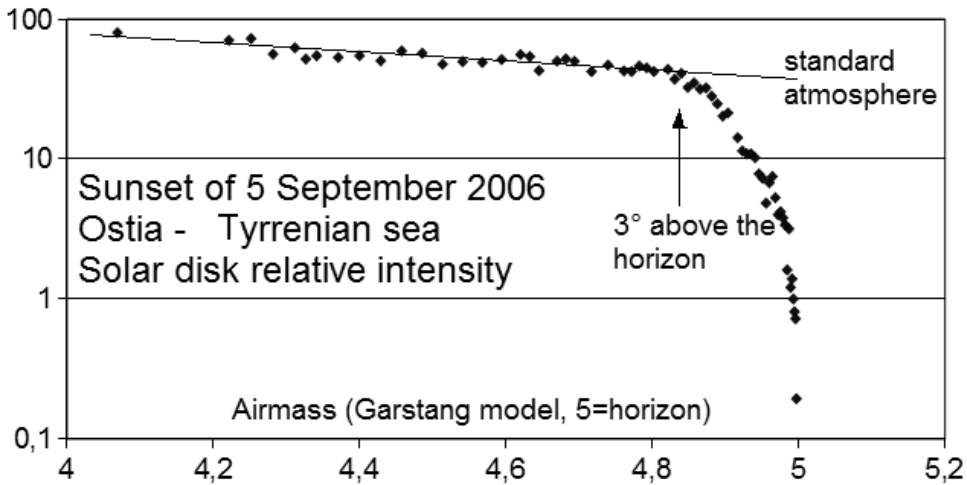,width=13cm}}
\vspace*{0pt}
\caption{Sunset of September 5, 2006: the altitude of the Sun above the horizon is measured in standard airmasses, according to the model of Garstang. 5 airmasses correspond to the horizon. A departure from the standard power law under the standard atmosphere, of the luminosity of the solar disk, starts from 3 degrees above the horizon.}
\end{figure}

A local flat, thin shield of humidity, with a few meters of scale height, explains that feature of the atmospheric transmission,  superimposed to the 5 airmasses of standard atmosphere,\cite{Garstang} right on the line of sight of the horizon.   

When the Sun approaches the horizon the connection between the image of the Sun and the horizon is anticipated by white filaments. Their occurrence is different day by day, it depends on the meteorology. During some occasions the first contacts occurred between the Sun and their reflected image on an atmospheric layer which is higher than the horizon.

For our method this layer is considered constant during the all time of the sunset. 

But how the black drop effect affects the timing of the first contact?
Indeed the general effect is to anticipate the time of the contact, and the larger is the instrument used for the observation the smaller is the anticipation.

To avoid this effect, if a video is available, it is possible to fit the profile of the Sun with a circle and the horizon with a line, in order to extrapolate the time of contact with its experimental uncertainty (dependant on the seeing along the line of sight).

\subsection{Rogue waves}

The second contact is much less dependant on the conditions along the line of sight.
The last glimpse of light can be affected only by the passage of a sea wave along the line of sight. During our sunsets the horizon was farther than 10 Km and the waves $w \sim 0.1$m, consequently an angular error on diameter reduction $\le 2$ arcsec can be possible, because of the anticipation of the last light by the occulting wave.
Nevertheless the accuracy on the timing is the main concern for this method. Videos with 60 frames per second allow a nominal accuracy of about 0.25 arcsec, and in this case the occulting wave can be influent, but in the case of the first contact the errorbar on the time is much larger.
The possibility to observe the Sun through the bottom of a wave is not excluded: along the line of sight there are other higher waves but the perspective can limitate their effect.
Using a video it is possible to set a reference horizon, despite of the moving waves, overcoming completely this problem.

\subsection{Parallax}

The point where the Sun touches firstly the line of the horizon is different from the one where the last light of the Sun disappears.
If the instrument is left in the same position during all the sunset there is no error of parallax, otherwise it can be introduced.
With an horizon at 10 Km a motion of 1 cm, to center the position of the Sun in the field of view, corresponds to 0.2 arcsec.
Moreover the difference of azimut between the first and the second contacts corresponds, at 42 degrees of latitude and with an horizon at 10 Km, to about 100 m. The constancy of the physical conditions of the atmosphere on such an horizontal distance over the sea surface is the limit of this hypothesis. This limiting factor is also present in all astrolabes measurements, dealing with almucantarats at definite altitudes.
As in the case of sunsets, the passage of the Sun across an almucantarat does not occur perpendicularly to it, unless it would be at the Equator in the solstice days, therefore if the first contact occurs on-axis, the second contact will be slightly off-axis, introducing different optical errors in the measurement. These errors, not being the same at the first and the second contacts, do not cancel each other.

\subsection{Point-like vs extended-arc seeing}

Along the line of sight there are several phenomena of turbulence occuring  in the atmosphere. The paths of the Sun beams are scattered like in a random walk process whose average dispersion is proportional to $\sim\sigma\sqrt N$.
Let consider a single point on the Sun disk, like a solar spot or the western point on the solar limb. During a drift observation, with the telescope and the detector in a fixed position, this point is moving along a zig-zag path. The standard deviation of the points of the path with respect of a line is a measurement of the point-like seeing at the moment of the observation.
Let change the angular scale of the analysis, by fitting the visible part of the solar limb (e.g. 10 arcmin) with an arc of a circle, or with a parabola. In this case the dispersion of the positions of the vertex of the parabola is much smaller than the point-like situation, because these positions are the result of an average of several individual points. In the case of automated methods in use with the modern heliometer in use at the Observatorio Nacional do Rio de Janeiro,\cite{helio} or DORAYSOL-type astrolabes\cite{Laclare} (Calern-France and Rio de Janeiro) or drift-scan observations at bigger telescopes (Clavius project\cite{clavius08} in Locarno-IRSOL and in Paris-Carte du Ciel) there are 100-200 points.
For the Solar Disk Sextant\cite{Sofia} only 5 points are used, but the telescope is operating at the top of the troposphere at 37 km of height, with only 3 mbar of residual pressure, with practically no residual turbulence along the line of sight.
At an altitude corresponding to 1.5 airmasses, where airmass(z)=1/cos(z), i.e. about 48 degrees of zenital distance, 30 degrees above the horizon, typical values of the point-like seeing are about 2-3 arcsec if observed from ground (Paris, Locarno, Roma, Rio de Janeiro).
The extended-arc seeing is as good as 0.6 arcsec, because of the statistical improvement of the factor $\sim \sqrt 100\div \sqrt 200$.
There is at least a factor of 2 missing, and this can be attributed to the different paths feasible by each solar ray coming from each point of the limb through the optical defects of the atmosphere generated by the turbulent cells along the line of sight.

Summarizing the situation is the following: 
\begin{itemize}
\item{point-like seeing $\sim\sigma$= 2 arcsec;}

\item{dispersion for each point of the limb observed due to N random walk process $\sim\sigma\sqrt N$;}

\item{statistical average to obtain the fitted position of the vertex over 200 points: $\sim\sigma\sqrt N/\sqrt(200)$=0.6 arcsec; the number of random scatters for 1.5 airmasses is $N\sim 18$.}

\item{For the sunset where the airmass number is 5 the number of random scatters increases proportionally $N\sim 60$, and the extended-arc seeing becomes 2 arcsec, while the point-like one is expected to be 6.6 arcsec.}

\item{Instruments with aperture smaller than 10 cm are diffraction limited. Through small aperture the modulation of the seeing are always amplificated to the diffraction limit. E.g. with a pinhole of 1.6 cm, point-like seeing of 5.4 arcsec are always measured,\cite{Sigi06} and this is the theoretical diffraction limit for the pinhole, and not the actual point-like seeing at the moment of the observation.} 
\end{itemize}
\subsection{Ephemerides} 
In the method of fixed almucantarats their altitude is determined by calculating it with the solar ephemerides at the time of the contact. What it is necessary is the constancy of the prisms angle of the astrolabe during the transit.
The sea horizon is indeed constant during a sunset, but it's depression depends on the height of the observer above the sea level and on the stratification of the atmosphere along the line of sight.
The absolute timing of the contacts allow to recover the position of the Sun in the celestial sphere, and the real depression of the horizon which can variate significantly with respect to the standard average value of 35 arcmin. 
The height of the almucantarat over which the transit occurs does not matter in the calculus of the diameter, because is the azimut of the Sun to be involved, but in any case the ephemerides play a role in the calculus for the accuracy of the azimut.
Generally speaking the ephemerides cannot affect the measurement for more than 0.01 arcsec.

\section{The Formula and Results}

\noindent From the spherical astronomy there is the formula\cite{Green}  

$dz/dt = 15 \times cos(lat) \times cos(azimut)$ arcsec/s.

\noindent This formula describes the variation of the zenital distance as function of the local latitude, and the solar azimut at a given time.
I have applied this formula for the value of the azimut at the instant $h$ in which the center of the Sun lies on the horizon, i. e. at the middle of the sunset.
For the calculation of the vertical solar diameter I used the approximated formula
$\Delta z = 15 \times cos(lat) \times cos(azimut_h) \times \Delta t$ arcsec
which works reasonably well with the astrolabes data (accuracy within $\pm 3$ arcsec verified on 24 February 2011 at the astrolabe of Rio de Janeiro).
A sunset observed in Ostia at latitude $41.73^o$ on February 2nd 2011 with naked eye at $azimut_h=247.88^o$, lasted $ \Delta t = 190 \pm 1 $ s and the corresponding calculated diameter is $\Delta z = 1970 \pm 10$ arcsec, with the ephemeris value of 1947.6 arcsec.
The video of August 11, 2009 showed a sunset in Ostia at latitude $41.74^o$ and $azimut_h=290.92^o$ lasted $ \Delta t = 178.4 \pm 0.1 $ s and the corresponding calculated diameter is $\Delta z = 1865.2 \pm 1.0$ arcsec, with the ephemeris value of 1894 arcsec.

\section{Conclusions}

The sunset is the natural phenomenon in which it is possible to study and timing the transit of the Sun over a natural reference line: the sea horizon. By analogy with the method of the solar astrolabe, the altitude of the sea horizon is not measured; only the absolute times of the contacts of the solar image (upper and lower limb) with that almucantarat are measured. In the computation of the solar diameter the azimut of the Sun at the moment of the observation is calculated with the ephemerides; the function of the astrolabe is to define an accurately horizontal 
almucantarat, that thing for the sunsets is guaranteed by the sea horizon.    

The initial idea was to demonstrate the possibility to measure by the naked eye the diameter with an accuracy below one arcmin, which is the diffraction limit of the eye. The measurements are rather complex, also with video, because the images are often saturated in the upper part (see Fig. 3).

\begin{figure}
\centerline{\psfig{file=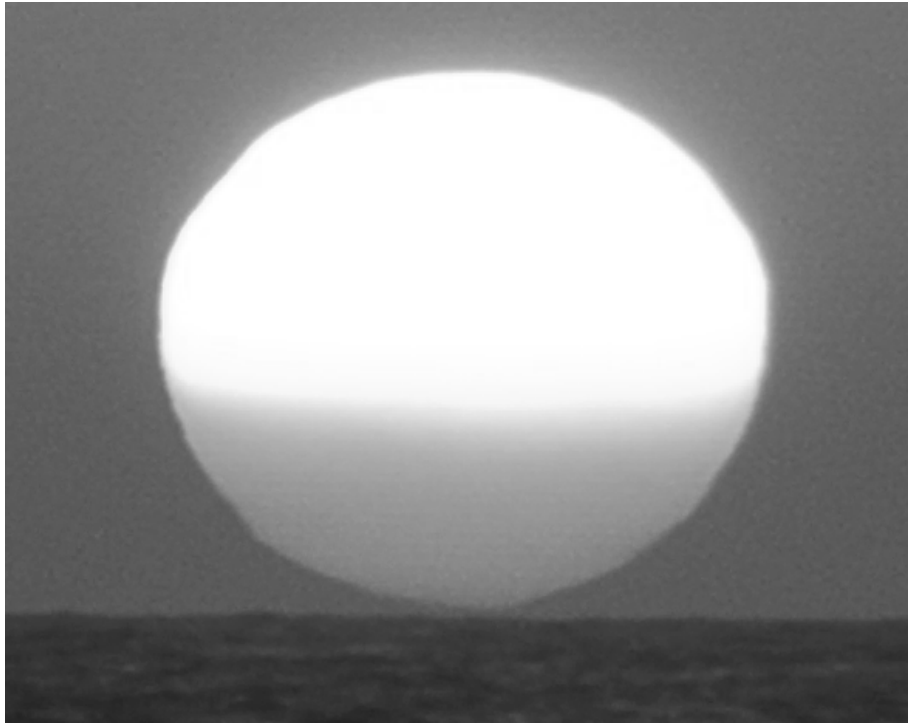,width=12cm}}
\vspace*{0pt}
\caption{Sunset of August 11, 2009 in gray scale. The upper part of the solar disk is full saturated, while the lower is right above the background.}
\end{figure}

It is very difficult to find a suitable filter which exploits the dynamical range of the video camera, as well as for the eye which is bleached by the Sun and cannot see easily the first contact.

This method allows to evaluate all the problems related with such kind of measurements; the most relevant from the optical point of view is the parallax error. Along the line of sight it is required that the atmosphere is stable during the whole phenomenon (up to 3 minutes for sunsets at Mediterranean latitudes and even 6 minutes for transits at the astrolabes) and it is homogeneous over about half a degree which is the difference in azimut between first and last contact at Mediterranean latitudes. 
In the case of the  astrolabe of Rio de Janeiro the problem due to the difference between first and second contact is reduced to 18 arcmin by the fact that the Sun moves more perpendicularly with respect to the almucantarats, when it is far from the meridian transit.
This parallax error and the lack of stability of the atmosphere during the minutes of the transit can be eliminated only by the repetition of the measurements by statistical average.

The observations of sunsets are extraordinary occasions for the didactic of solar astrometry methods.

\end{document}